\documentstyle[12pt,subfigure,glas]{article}
\begin{document}
\begin{titlepage}{GLAS--PPE/95--04}{\today}
\title{GaAs Microstrip Testbeam Results}


\begin{Authlist}
R.L.~Bates\Instref{phy_gla}
C.~Buttar\Instref{phy_shef}
S.~D'Auria\Instref{phy_gla}
C.~del Papa\Instref{udine}
W.~Dulinski\Instref{lepsi}
S.J.~Gowdy\Instref{phy_gla}
S.~Manolopoulos\Instref{phy_shef}
V.~O'Shea\Instref{phy_gla}
T.~Sloan\Instref{phy_lan}
C.~Raine\Instref{phy_gla}
K.M.~Smith\Instref{phy_gla}
F.K.~Thomson\Instref{phy_gla}
\end{Authlist}

\Instfoot{phy_gla}{Dept. of Physics \& Astronomy, University of Glasgow, UK}
\Instfoot{phy_lan}{Dept. of Physics, University of Lancaster, UK}
\Instfoot{phy_shef}{Dept. of Physics, University of Sheffield, UK}
\Instfoot{udine}{Dip. di Fisica, Universit\`a di Udine, Italy}
\Instfoot{lepsi}{L.E.P.S.I., Strasbourg, France}

\collaboration{On Behalf of the RD8 Collaboration}





\begin{abstract}
A gallium arsenide detector was tested with a beam of 70GeV pions
at the {\sf SPS} at CERN. The
detector utilises a novel biasing scheme which has been shown to behave as
expected. The detector has a pitch of 50$\mu$m and therefore an expected
resolution of 14.5$\mu$m. The measured resolution was approximately
14$\mu$m. By using
a non-linear charge division algorithm this can be increased to $\approx$
12$\mu$m. Noise was the limiting factor to the resolution. This was 2000e$^-$
as opposed to the expected 360e$^-$. This noise is also thought to have
reduced the detection efficiency of the detector. The source of the excess
noise is currently being investigated.
\end{abstract}

\vspace{2cm}

\begin{center}
\em Presented by S.J.~Gowdy at the \\
$4^{th}$ workshop on GaAs detectors and
related compounds \\
San Miniato (Italy) 19-21 March 1995
\end{center}


\end{titlepage}

\section{Introduction}

As part of the ongoing programme developing radiation hard detectors
for the LHC a test beam run was carried out in September 1994. 

\section{Testbeam Setup}

\begin{figure}[b]
\centerline{\epsfig{file=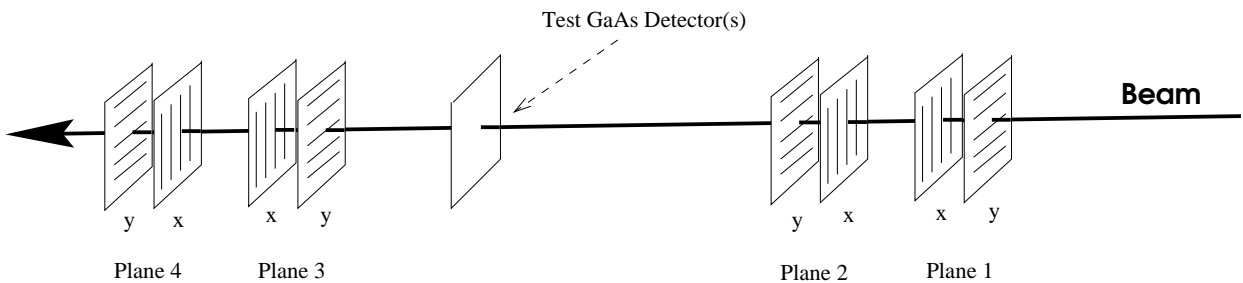,width=\the\textwidth}}
\caption{Strasbourg Telescope Layout\label{fig:tele}}
\end{figure}

The X1 beam from the {\sf SPS} at {\sf CERN} was incident on a
telescope provided by L.E.P.S.I., Strasbourg. This beam
was a tertiary  beam: pions with an energy of
70GeV were used for the test.
This telescope employs eight silicon microstrip detectors and five
trigger scintillators.
The silicon detectors
had a 25$\mu$m implant
pitch, but strips were only read-out every 50$\mu$m. The intermediate
implant strip produces enhanced resolution between the strips. Each of
these detectors was 300$\mu$m thick.
A schematic representation of the telescope can be seen in figure
\ref{fig:tele}.

The silicon detectors and the test detector were read out with a Viking
pre-amplifier chip\cite{tok:93}. This produces a multiplexed signal which is
sampled with a Sirocco ADC and then written to an {\sf EXABYTE} tape. The
Siroccos and the tape drive are housed in a {\sf VME} Data Acquisition
({\sf DAQ}) system. This allowed limited on-line analysis.

For an event to be written to tape it had to satisfy the trigger
conditions,
namely that the event occurred during the beam extraction phase of the
{\sf SPS}, the {\sf DAQ} was not {\sf BUSY} and finally, and most
importantly,
there were coincident hits in a number of the scintillators.

\section{Detector Structure}

\begin{figure}
\begin{center}
\begin{tabular}{cc}
\subfigure[Simplified plan of detector]%
{\epsfig{file=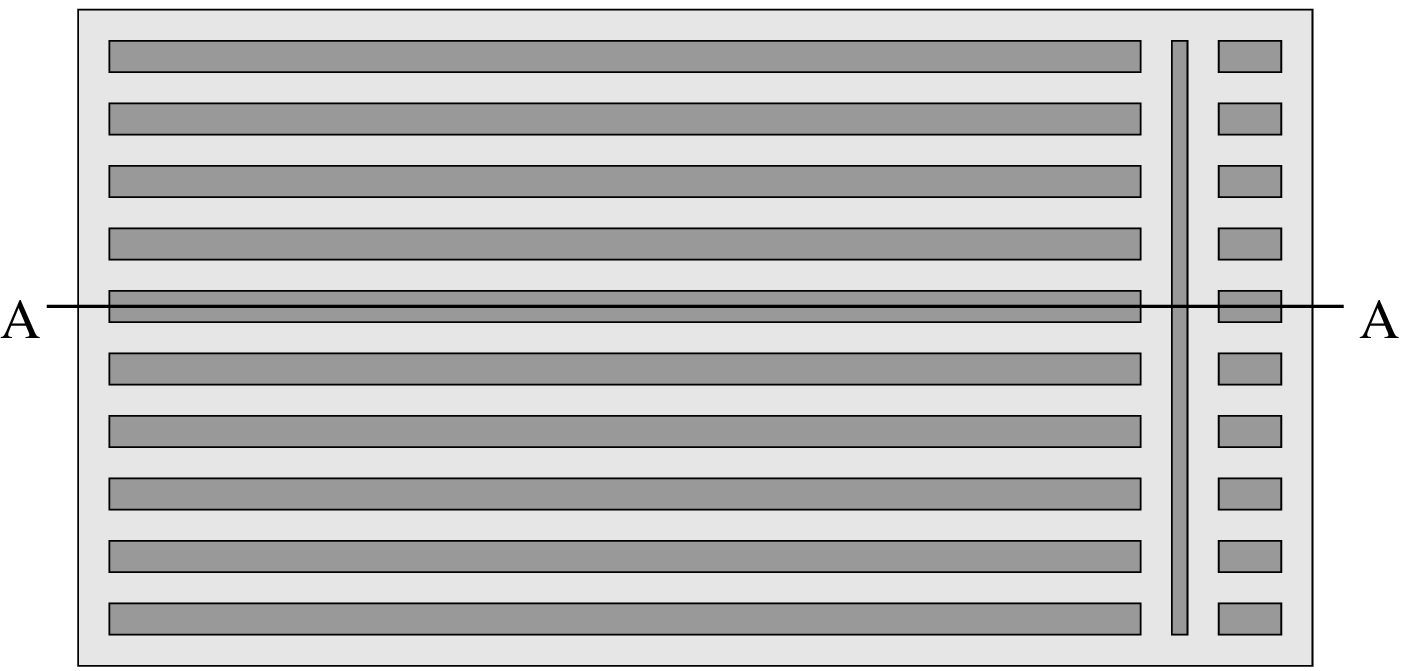,width=.45\textwidth}}
&\subfigure[Cross section of detector]%
{\epsfig{file=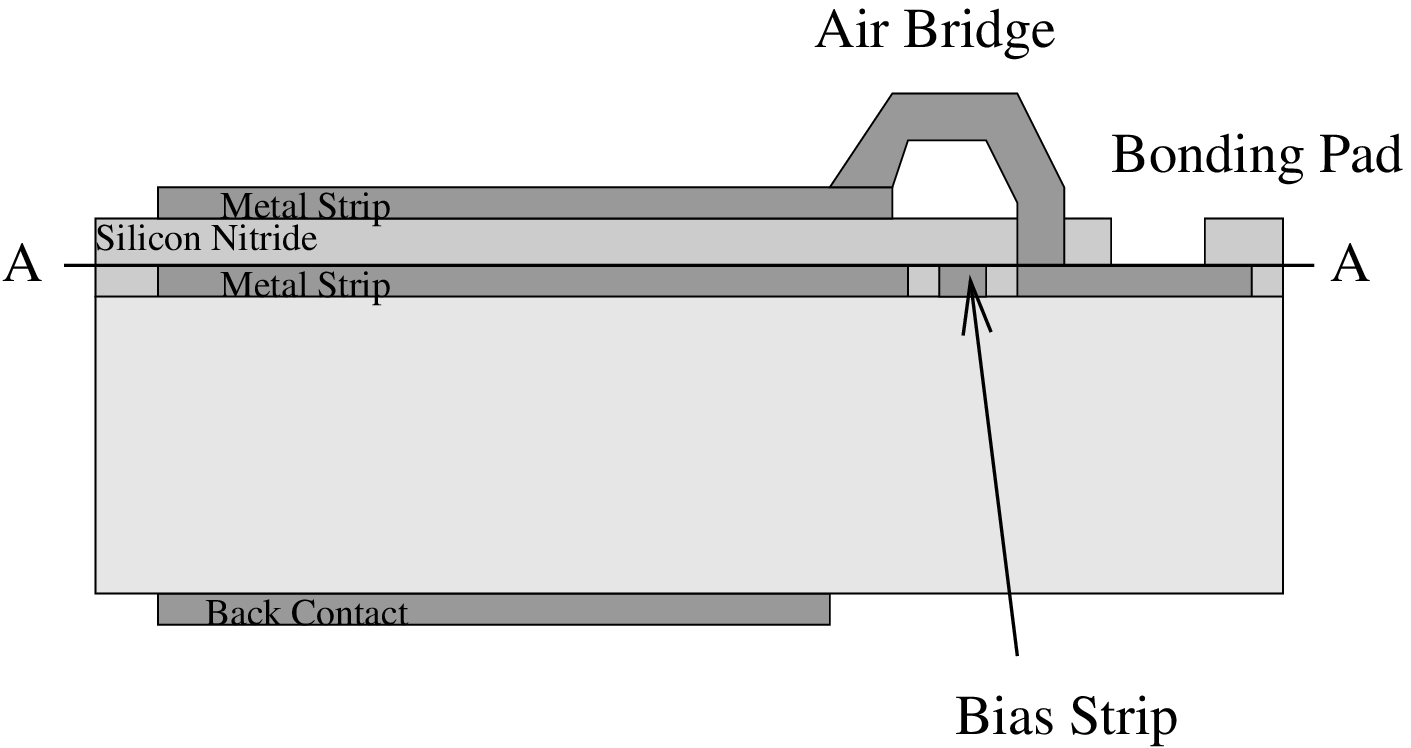,width=.45\textwidth}}\\
\end{tabular}
\end{center}
\caption{Representation of an Alenia AC-coupled detector\label{fig:xsec}}
\end{figure}

The gallium arsenide detector tested was fabricated by Alenia SpA,
Rome. It had both 50$\mu$m pitch and readout. The detector had an integrated
Si$_3$N$_4$ capacitor and novel biasing
scheme to keep the detector strips close to
zero volts. The capacitor was between two layers of metal, the lower was
in contact with the substrate and was 30$\mu$m wide and the upper was 20$\mu$m
wide. The cross-section of a detector and
an illustration of the top of the detector are shown in figure \ref{fig:xsec}.

The gap between the biasing strip and the detector strip was 5$\mu$m long and
6$\mu$m wide. The thickness of the detector was 200$\mu$m.

\section{Detector Characteristics}

\begin{figure}
\centerline{\epsfig{file=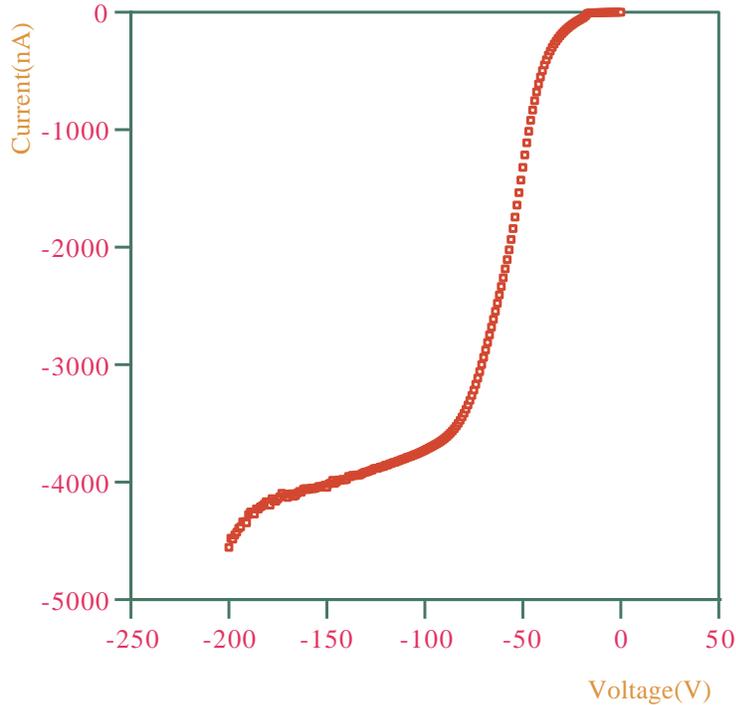,width=.6\textwidth}}
\caption{Current-Voltage Characteristics of detector\label{fig:iv}}
\end{figure}

Before the detectors are tested in a beam they are characterised
electrically. The basic test for a detector is to verify that it
operates as a reverse-biased diode with acceptable current.
The I-V characteristic is shown 
for the detector in figure \ref{fig:iv}.

\begin{figure}
\begin{center}
\begin{tabular}{c}
\subfigure[Schematic of VV Circuit]%
{\epsfig{file=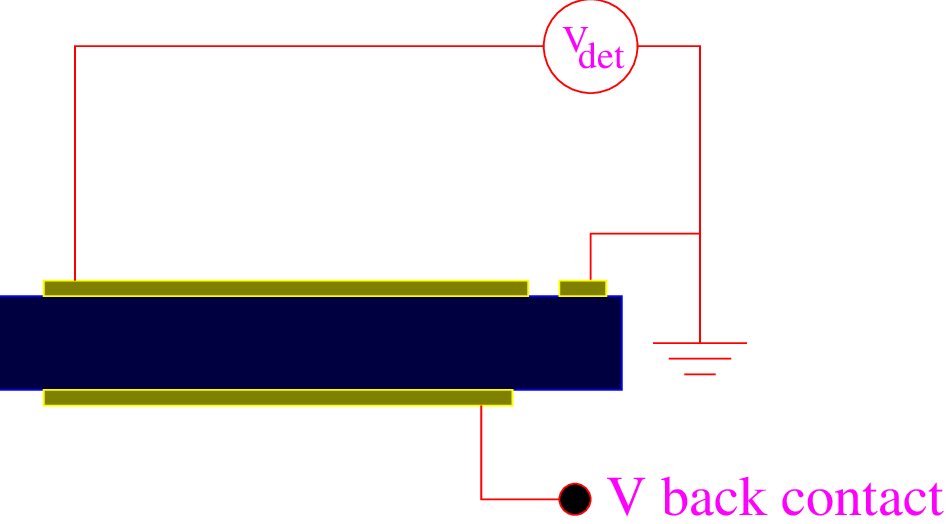,height=.6\textwidth,angle=270}
\label{fig:vv-cir}} \\
\subfigure[Voltage on detector strip vs voltage on back contact]%
{\epsfig{file=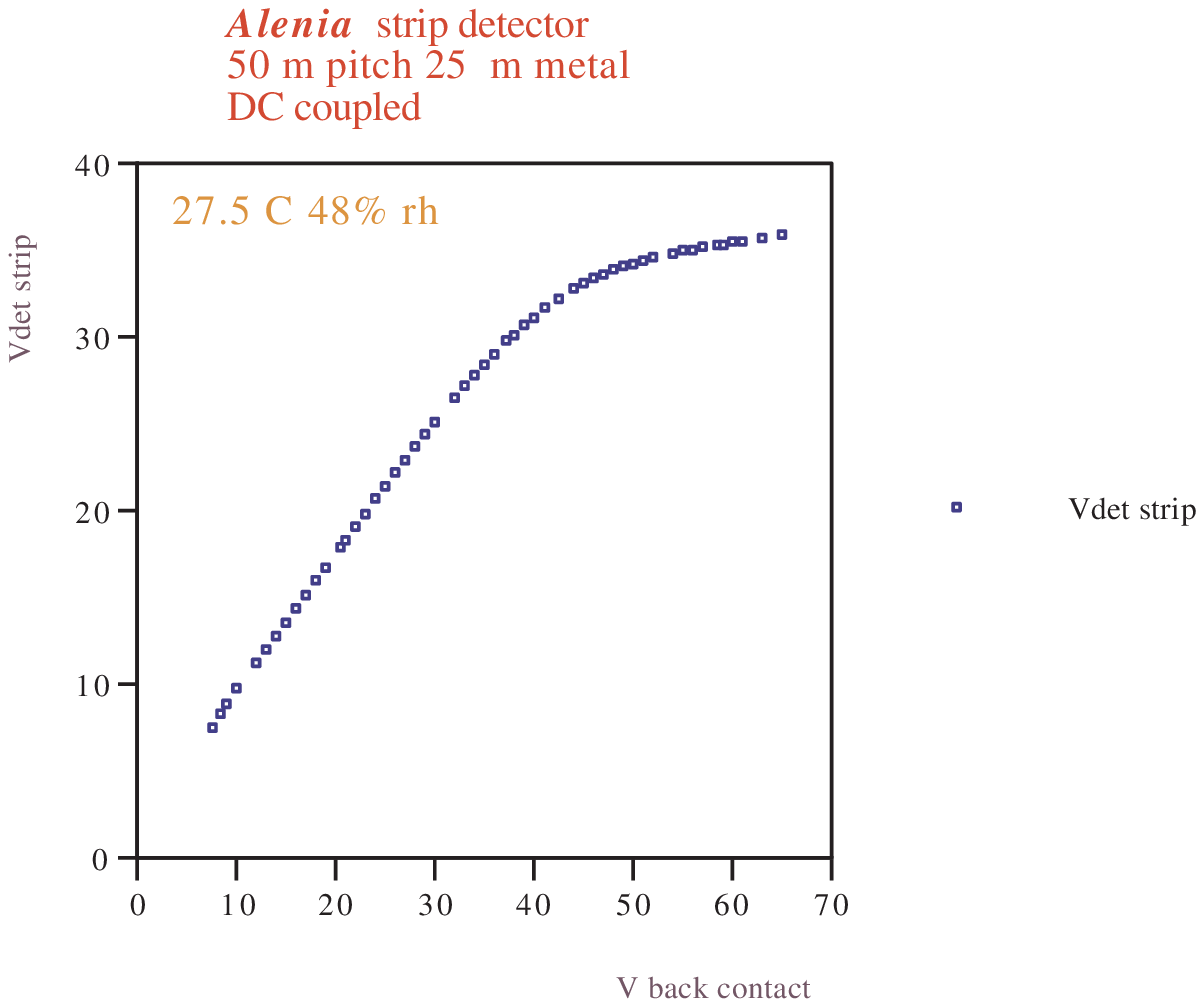,width=.8\textwidth}\label{fig:vv}} \\
\end{tabular}
\end{center}
\caption{Voltage-Voltage Figures}
\end{figure}

To test the novel biasing structure the voltage on the detector strip
was plotted against the voltage applied to the rear contact, using
the circuit shown in figure \ref{fig:vv-cir}. 

The results of the test show that with the biasing strip held at zero
volts the detector strip will float up to about 35 volts. Therefore,
during the testbeam this strip was held at -30V. The results are
shown in figure \ref{fig:vv}.

\section{Off-line Analysis}

The off-line analysis consisted of two separate stages. The first of these
read in the {\sf RAW} data. The first 100 events are used to determine the
pedestal levels of the data and the next 100 to determine 
the noise. This program also
allows for a group of pedestals shifting coherently, in what is known
as Common Mode Shift({\sf CMS}).

After the initialisation phase the noise is continuously updated,
with a weight
of 50 on the old value to reduce the effect of actual hits which
fall below the $\frac{S}{N}$ cut on the noise. It then looks through the
remaining data for clusters.

A cluster is found if the $\frac{S}{N}$ value for any strip is above 3(4).
A search is then made for any adjacent strips with $\frac{S}{N}$
above 1.5(2). If found,
these determine the width of the cluster for GaAs(Si);
the cluster total signal and noise are then calculated
using equations \ref{equ:sig} and \ref{equ:noi}.

\begin{equation}
S=\Sigma S_{strip}
\label{equ:sig}
\end{equation}

\begin{equation}
N=\sqrt{\frac{\Sigma N_{strip}^2}{\# strips}}
\label{equ:noi}
\end{equation}

If the cluster is wider than a single strip, its $\eta$ value 
is evaluated using equation \ref{equ:eta}. This is used to enhance the
resolution using a non-linear charge sharing algorithm. 

\begin{equation}
\eta=\frac{S_L}{S_L+S_R}
\label{equ:eta}
\end{equation}

\begin{figure}
\begin{center}
\begin{tabular}{cc}
\subfigure[$\eta$ distribution]%
{\epsfig{file=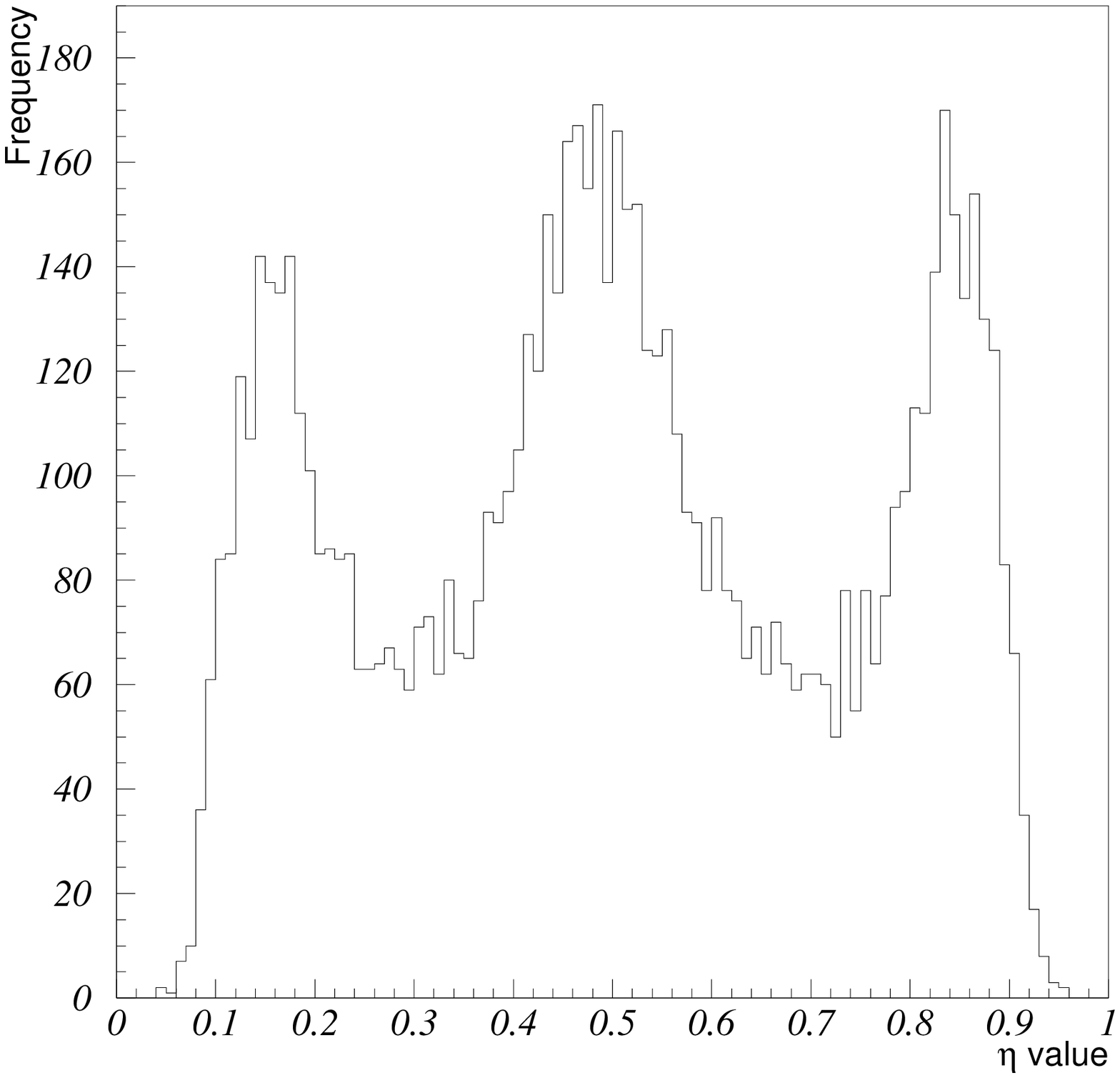,width=.45\textwidth}\label{fig:eta}}
&\subfigure[Integrated $\eta$ distribution]%
{\epsfig{file=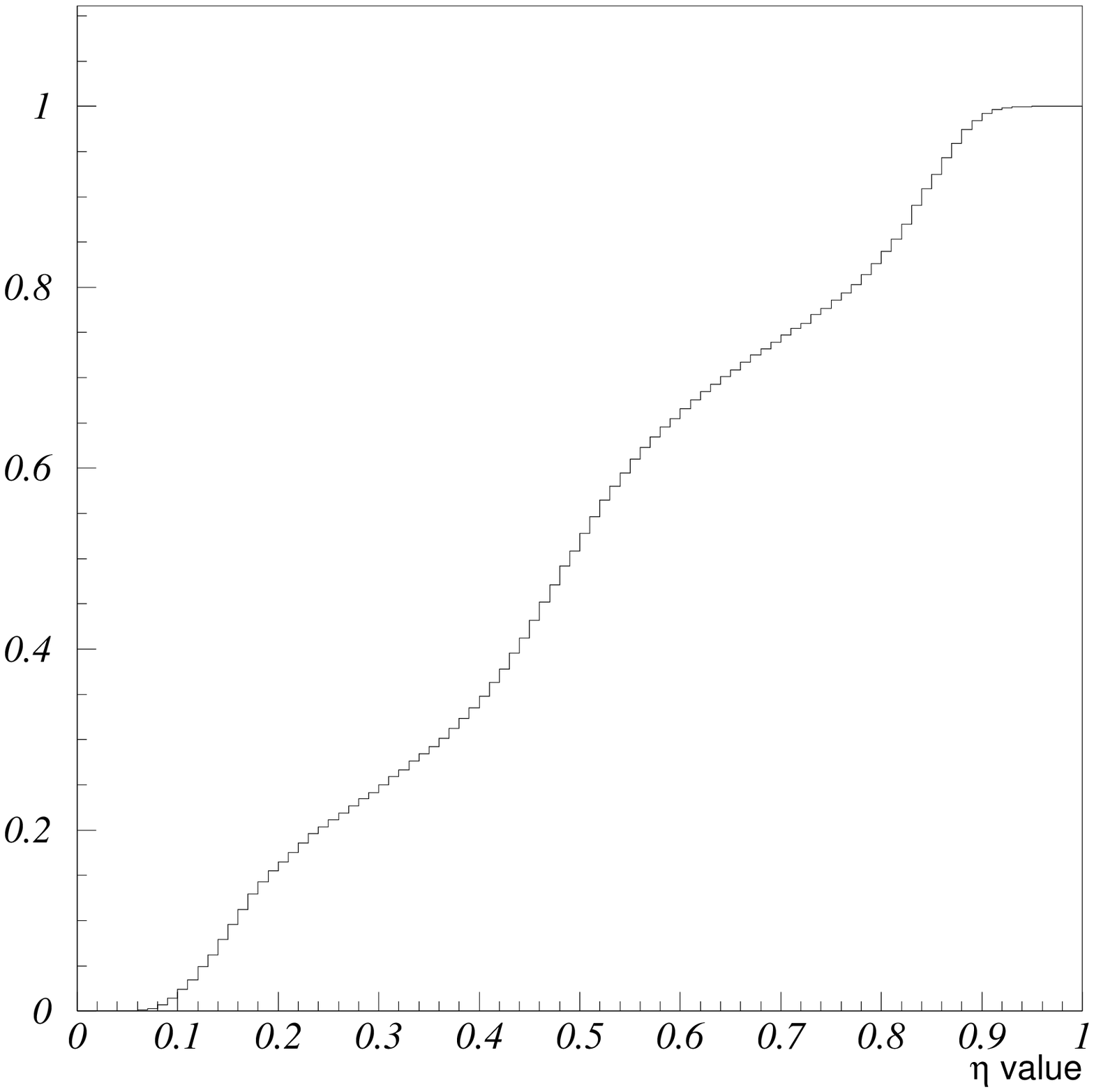,width=.45\textwidth}\label{fig:inteta}}
\end{tabular}
\end{center}
\caption{$\eta$ distributions in a silicon detector}
\end{figure}

In this equation $S_L$($S_R$) is the signal on the left(right)
of the the two maximum pulse height strips of a cluster. A typical
$ \eta $ distribution from a silicon detector is shown in figure \ref{fig:eta}.

By integrating the $\eta$ distribution we define a look-up table for
the position of a hit between the two strips. The integrated distribution is
shown in figure \ref{fig:inteta}.

The cluster information is then written to a {\sf DST} file together with the
pulse height and noise
values for the five strips on either side of the cluster centre.

This file is then read in by the track fitting program. It uses the hits
in the {\sf DST} file to determine the relative alignment of the
detectors. First, the telescope is
self-aligned with its origin at the centre of the first two detectors. 

\begin{figure}
\begin{center}
\begin{tabular}{cc}
\subfigure[Initial Residuals vs distance along strip]%
{\epsfig{file=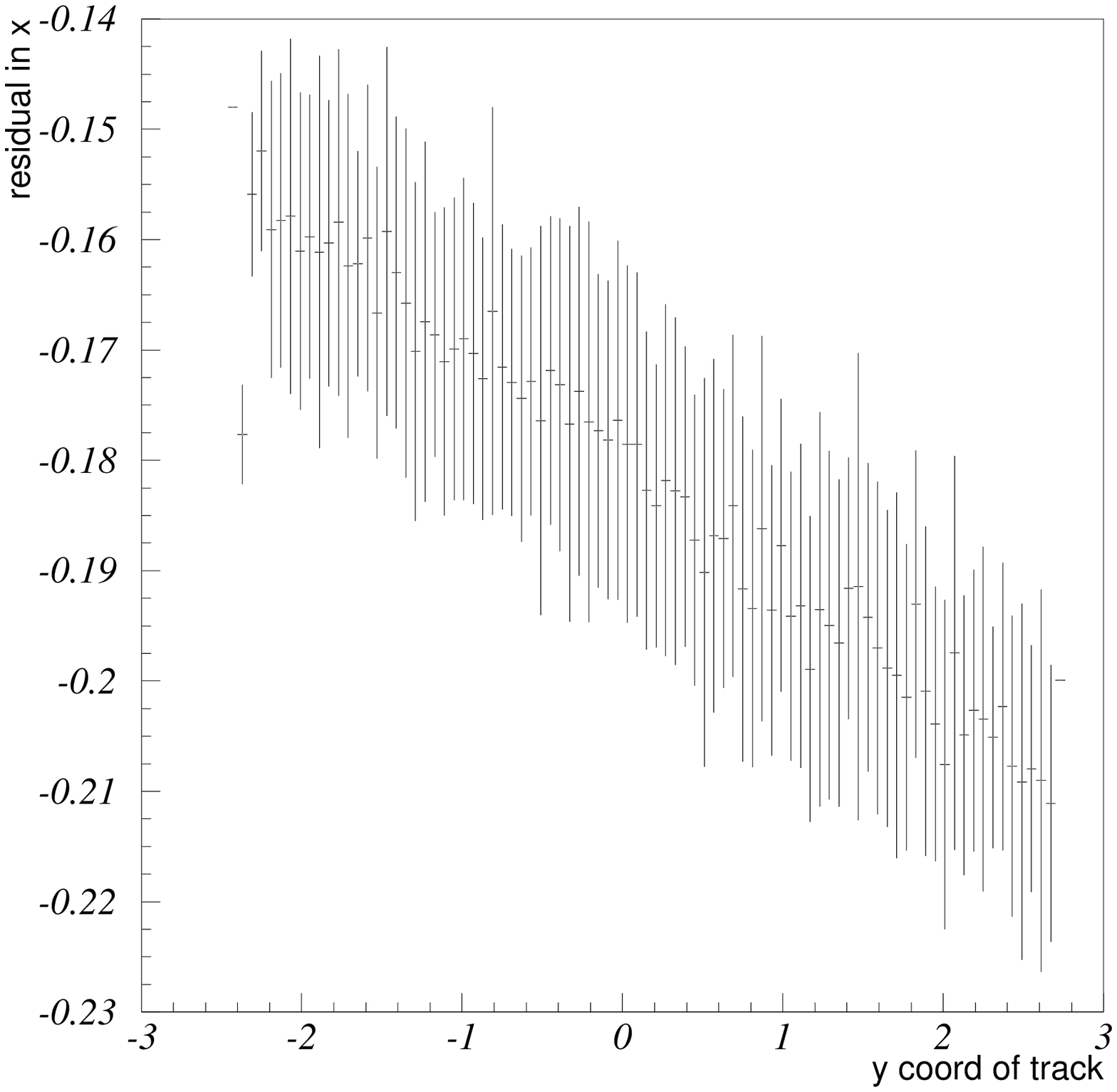,width=.45\textwidth}\label{fig:resi-y}}
&\subfigure[Resolution against distance along strip]%
{\epsfig{file=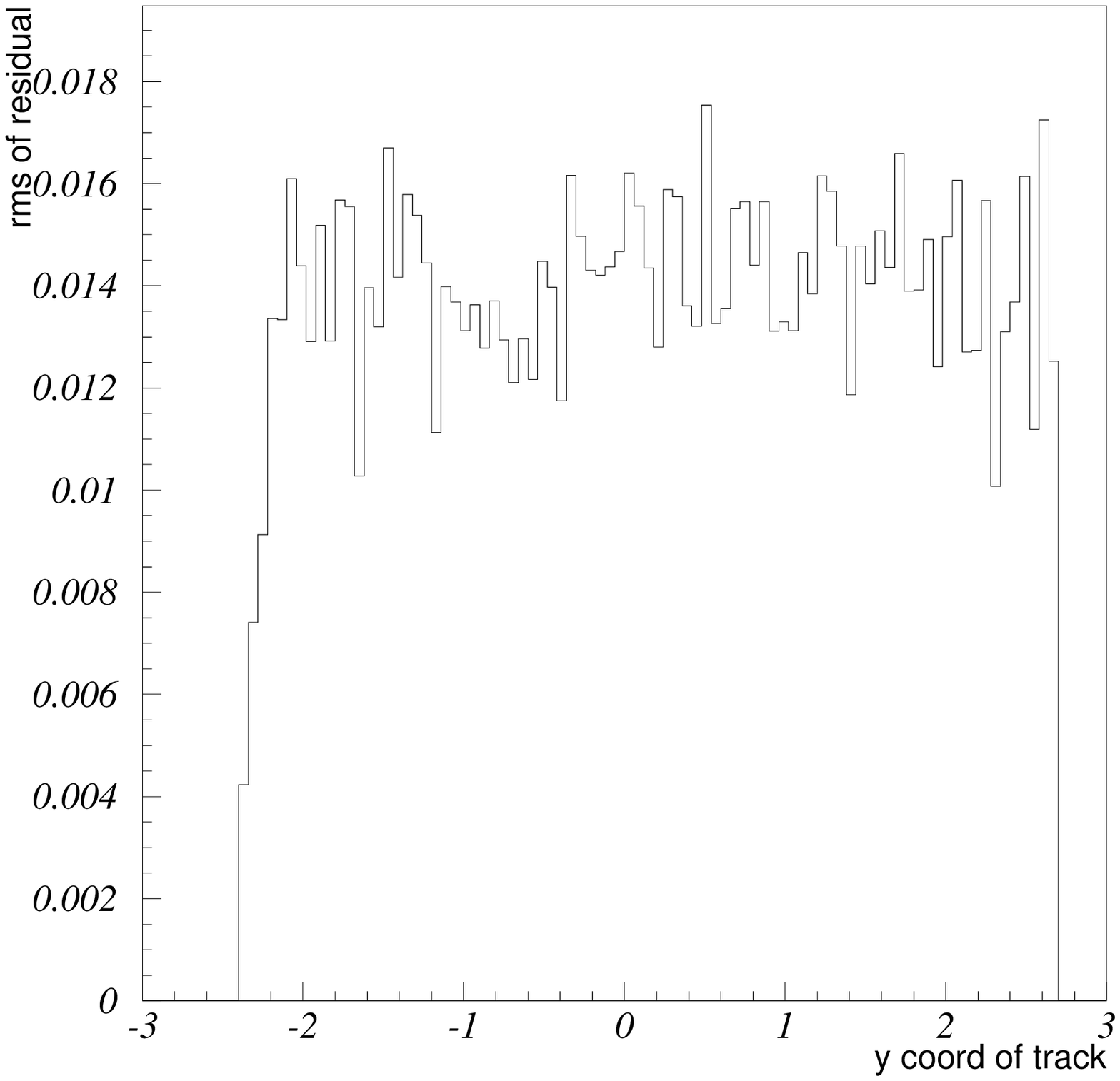,width=.45\textwidth}\label{fig:res-y}}
\end{tabular}
\end{center}
\caption{Variations along strip}
\end{figure}

Next the test detectors are aligned in this reference frame. The
first stage is carried out by centering the residual plot for the detector around
zero. Then, to allow for the strips not being orthogonal to the reference
frame, the residual in the detector is plotted against the position along
the strip. This plot is shown in figure \ref{fig:resi-y}. A linear 
fit is made to the data and the hits rotated to align the strips with the
reference frame.

\begin{figure}
\centerline{\epsfig{file=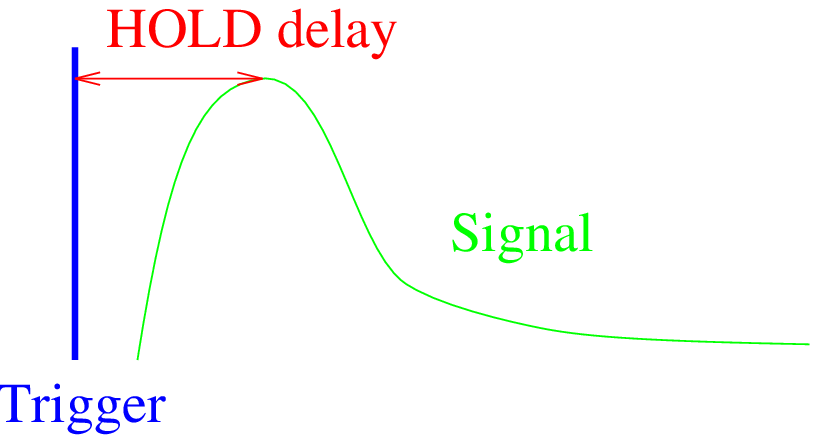,height=.4\textwidth,angle=270}}
\caption{{\sf HOLD} delay\label{fig:hold}}
\end{figure}

After this is complete the detector position along the beam line is adjusted
so that the fit residuals are minimised. The detector is then rotated
by a small angle
about the centre strip axis and the residuals are minimised again.

\section{Results}

The shaping time for the gallium arsenide was reduced from the standard
silicon time of 2100ns to 680ns to reduce the noise as the parallel noise
term is expected to dominate with gallium arsenide. This is implemented
by a potentiometer on the repeater card. On the {\sf DAQ}
this shaping time is stored as the {\sf HOLD} delay and
is the time from receiving a trigger to the expected peak in the
signal,
as illustrated in figure \ref{fig:hold}. During the test beam run 
there can be only one value for all
detectors: this number was
varied between 1000ns and 680ns to optimise the readout for the
gallium arsenide detector under test. Some degradation of
the silicon performance was accepted due to the reduced shaping time.

\begin{figure}
\begin{center}
\begin{tabular}{cc}
\subfigure[Landau Distribution]%
{\epsfig{file=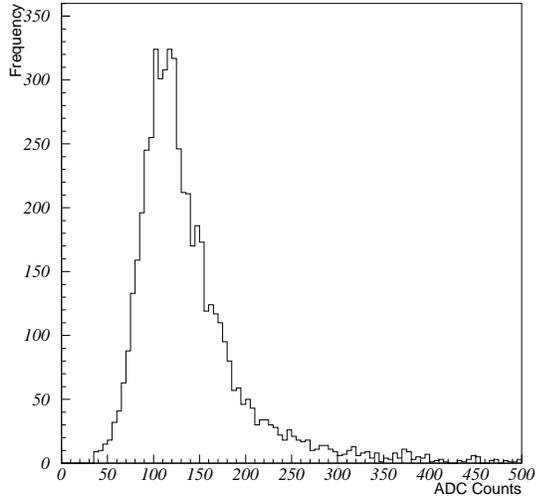,width=.45\textwidth}\label{fig:lan}}
&\subfigure[$\frac{S}{N}$ Distribution]%
{\epsfig{file=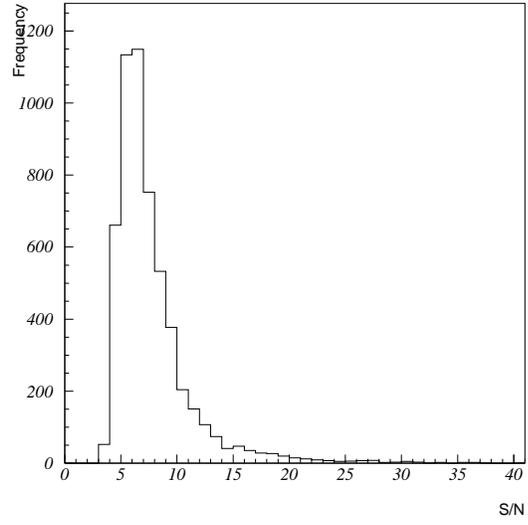,width=.45\textwidth}\label{fig:s-n}} \\
\multicolumn{2}{c}{
\subfigure[Residual Distribution]%
{\epsfig{file=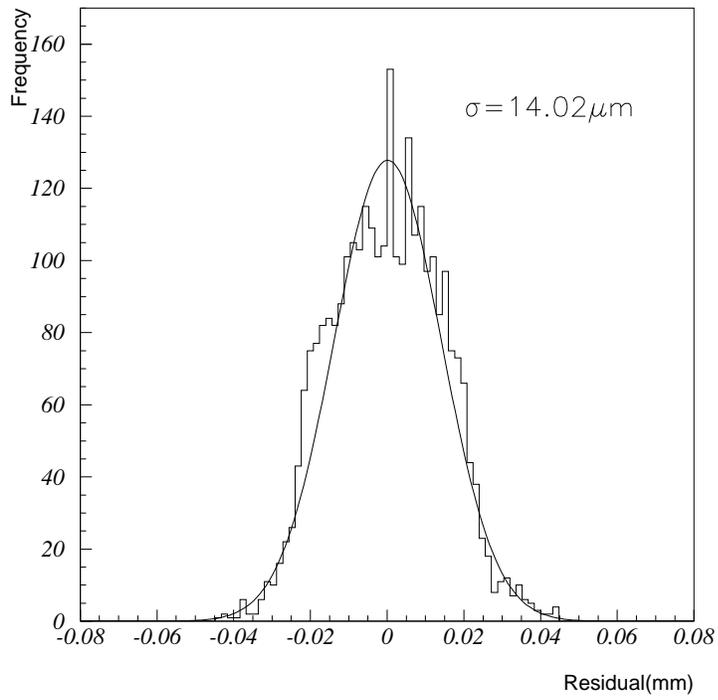,width=.6\textwidth}\label{fig:res}}}
\\
\end{tabular}
\end{center}
\caption{Results for detector AL-W3-6AC}
\end{figure}

The equivalent noise charge({\sf ENC}) was measured to be 2000e$^-$. The
signal-to-noise from the data
in all cases was very close to 6 (Figure \ref{fig:s-n}).
The signal was therefore 12000e$^-$,
which implies a charge collection efficiency({\sf CCE}) of 46\%.
This is in good
agreement with tests made on pad diodes which have 60\% {\sf CCE} but
were operating at 200V as opposed to 180V as used in the test beam.

The rms of the residual distribution shown in figure \ref{fig:res}
corresponds to a detector resolution of  $14.02\pm0.17 \mu$m. Figure
\ref{fig:res-y} shows that this is constant along the length of the
strips. This compares
well with the expected resolution from binary read out, that is without
charge-sharing information, as given by 
$\frac{pitch}{\sqrt{12}}=14.43\mu m $.

Using only clusters with a width of two or more strips, the
resolution improves to approximately 12$\mu$m.
It is observed that 50\% of events have clusters of
more than one strip, which is slightly more than expected from the gap
to metal width ratio, indicating a proportion are due to noise.

The detection efficiency at the shortest {\sf HOLD} delay time was 47\%.
This was the highest value obtained. Various investigations were carried out
on the data in an attempt to determine the
source of this poor performance. A search was made for inefficient
regions in both time and
space. However, none were found.

The low detection efficiency may be attributable to a problem
with the {\sf DAQ} which produced an asymmetric $\eta$
distribution for the silicon detectors which had been normal before this
test beam run. At present, however, it is not possible to prove that
this is the sole cause of the inefficiency.

\section{Simulation}

To investigate other possible reasons for the poor detection efficiency,
 a computer simulation
was done. The simulation program produced data in the {\sf RAW} format produced
by the {\sf DAQ} system, thus serving also as a test to validate
our use of the software.

In this simulation charge sharing is assumed to be linear in all
detectors. The charge deposited by a particle is then shared between
two strips. This charge is determined from a function
which produces a Landau distribution of signals. Once this
has been shared between the two strips Gaussian noise
is added to all strips.

\begin{figure}
\begin{center}
\begin{tabular}{cc}
\subfigure[Graph of $\frac{S}{N}$ and Detection Efficiency]%
{\epsfig{file=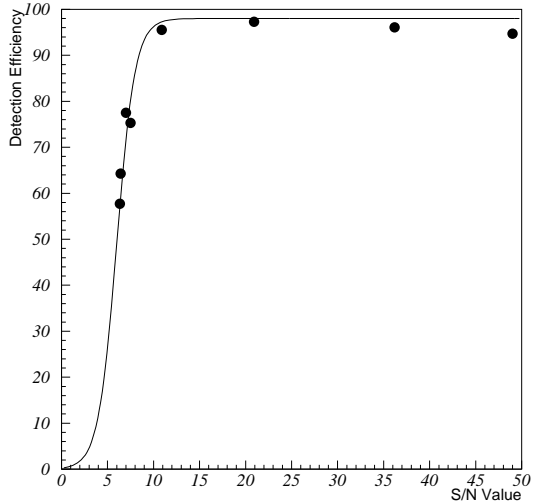,width=.45\textwidth}\label{fig:det-eff}}
&\subfigure[Distribution of hits not seen with respect to centre of strip]
{\epsfig{file=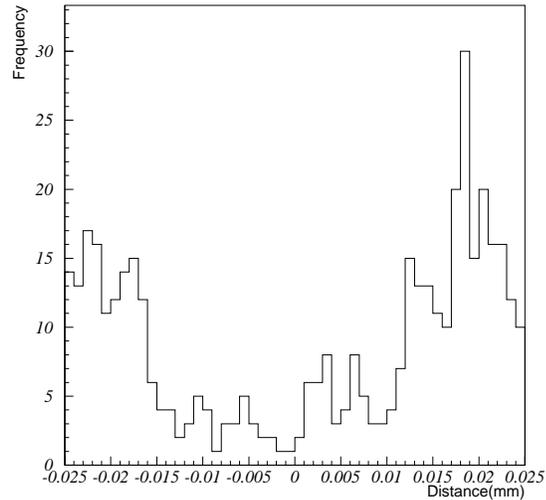,width=.45\textwidth}\label{fig:mis}}
\end{tabular}
\end{center}
\caption{Results of test beam simulation}
\end{figure}

The resulting $\frac{S}{N}$ is shown in figure \ref{fig:det-eff}. It
appears
that, for a $\frac{S}{N}$ value below approximately ten, the detection
efficiency drops dramatically.

This is due to the charge sharing dividing the signal
and a conspiracy of noise and Landau fluctuations pushing the signal
below the threshold of 3$\sigma_{(\frac{S}{N})}$.

This would occur more readily toward the centre of the gap between
the strips, as observed in the simulation (shown in figure
\ref{fig:mis}), however this was not seen in the actual data.

\section{Conclusion and Further Work}

The noise observed in the test beam data and with the detector
in the laboratory is over five times greater than expected (360e$^-$
{\sf ENC}\cite{tok:93}).
The source of this excess noise will be investigated further.

By reducing the noise both resolution and detection efficiency
will improve. The lack of resolution attainable with a given $\eta$ algorithm
is proportional to $\frac{N}{S}$ \cite{tur:93}. As
seen in figure \ref{fig:det-eff}, once the $\frac{S}{N}$ is
above ten detection efficiency should not be a problem.

At the LHC gallium arsenide detectors will be aided by the faster
shaping times of the pre-amplifiers, as 80\% of the signal will
be delivered within 20ns. 

Currently under investigation are non-minority carrier injecting ohmic 
contacts which will enable detectors to be more easily 
``over-depleted''. This will
increase the charge collection efficiency of the detectors and hence
the signal produced.

Studies to be made in 1995 include varying the gap/metal width
ratio, response at lower temperatures and keystone detector geometry.
In varying the gap/metal width the signal will decrease (due to poorer
field) but the resolution will increase. Since silicon detectors may
require an operating temperature of -10$^\circ$C
to prevent reverse annealing the gallium arsenide response at lower 
temperatures
needs to be checked. It is planned to use detectors with keystone
geometry in the ATLAS forward region so detectors must be fabricated in
this design to prove their viability. 

%

\section{Acknowledgements}

Off-line software was originally written by R. Turchetta, J.A. Hernando and 
J.A. Straver, to all of whom the authors are greatly
indebted. A. Rudge and R. Boulter also provided exceptional technical support.

\end{document}